\documentclass[9pt,conference]{IEEEtran}
\usepackage{amssymb,amsthm,amsmath,array}
\usepackage{graphicx}
\usepackage{xspace}
\usepackage[sort&compress, numbers]{natbib}
\usepackage{stmaryrd}
\usepackage{xcolor}
\usepackage{mathtools}
\usepackage{float}
\usepackage{textcomp}
\newtheorem{rem}{Remark}
\usepackage{subfigure}

\begin{document}
\title{Scatterer Localization Using Multi-Bounce Paths}
\author{\IEEEauthorblockN{
        Yuan Liu\IEEEauthorrefmark{1}, 
        Linlong Wu\IEEEauthorrefmark{1},  
        Xuesong Cai\IEEEauthorrefmark{2}\IEEEauthorrefmark{3}, and
        M.~R.~Bhavani Shankar\IEEEauthorrefmark{1}
    }
    \IEEEauthorblockA{
        \IEEEauthorrefmark{1} Interdisciplinary Centre for Security, Reliability and Trust (SnT), University of Luxembourg, L-1855, Luxembourg.\\
        \IEEEauthorrefmark{2} School of Electronics, Peking University, Beijing, 100871, China.\\
        \IEEEauthorrefmark{3} Department of Electrical and Information Technology, Lund University, 22100 Lund, Sweden.}
}

\maketitle
\begin{abstract}
    Indoor sensing is challenging because of the multi-bounce effect, spherical wavefront, and spatial nonstationarity (SNS) of the near-field effect. This paper addresses radio-based environment sensing considering these issues. 
Specifically, graph theory (GT) is used to model the multi-bounce propagation of the near field. In this manner, indoor reflectors/scatterers are modeled as vertices in a propagation graph, the multi-bounce paths are modeled by the edges linking the vertices. 
Besides, the coupled multipath parameters in the near field, i.e., range and angles, are denoted directly by the coordinates of vertices.  
Then, the space-alternating generalized expectation-maximization (SAGE) algorithm is adapted to the proposed Graph theory-based dictionary-aided Multi-bounce SAGE (GM-SAGE), where the searching parameters including range and angle of departure/arrival (AoD/AoA) are transformed to the coordinates of scatterers in the graph. 
The proposed algorithm is validated through measurement-calibrated ray tracing (RT) in a complex indoor office. 
The results demonstrate that the proposed GM-SAGE can deal with multi-bounce channels.
\end{abstract}

\section{Introduction}
\subsection{Near field} 
The Rayleigh distance, i.e., $ d_\text{Rayleigh} = 2 D^2 / \lambda$, defines the boundary of the near field and the far field \cite{BalanisConstantineA2005At:a}, where $D$ represents the aperture size of the array and $\lambda$ represents the wavelength.
As a result of using large-scale arrays to obtain high beamforming gains and high angular resolutions \cite{zhi2024performance,zhang2022scatterer,xing2021millimeter}, i.e., increasing of $D$, the near-field regions are enlarged. 
The incident angles from sources within the near-field region are not identical across the array.
The AoA and AoD of each array element are dependent on the range between the source and the exact element, and the array aperture, resulting in the so-called range-dependent angles.
This range angle spatial relationship of the near-field channel is modeled as a spherical wavefront \cite{10179246,10038714,liu20223d}.
Another observed feature of the near-field is the spatial non-stationarity (SNS) in the path gain across the array of elements \cite{8713575}.

\subsection{Multi-bounce sensing} 
Multi-bounce paths are commonly seen in wireless channels, especially in complex indoor environments \cite{ling2017experimental,de2023analysis}.
However, many existing works consider only one-bounce paths \cite{wang2021room}, or try mitigating the multi-bounce effects but not using them \cite{zhang20235g}. 
In a new trend of radio simultaneous localization and mapping (SLAM), which localizes unknown transmitter (Tx) and utilizes multipaths for environment sensing \cite{10556695}. In such applications, the multi-bounce paths, i.e., multipaths due to two or more reflections and scattering\footnote{Reflection is a special kind of wave scattering. In this abstract, scatterers generally include reflectors.} can provide rich environment information. 

This work presents the idea of utilizing multi-bounce paths of large-scale antennas for scatterer localization, which can act as a follow-up step after source localization in radio SLAM. 

\section{Problem formulation and solution}
%%%%%%%%%%
\begin{figure}[t]
    \centering
    \includegraphics[width=0.45\textwidth]{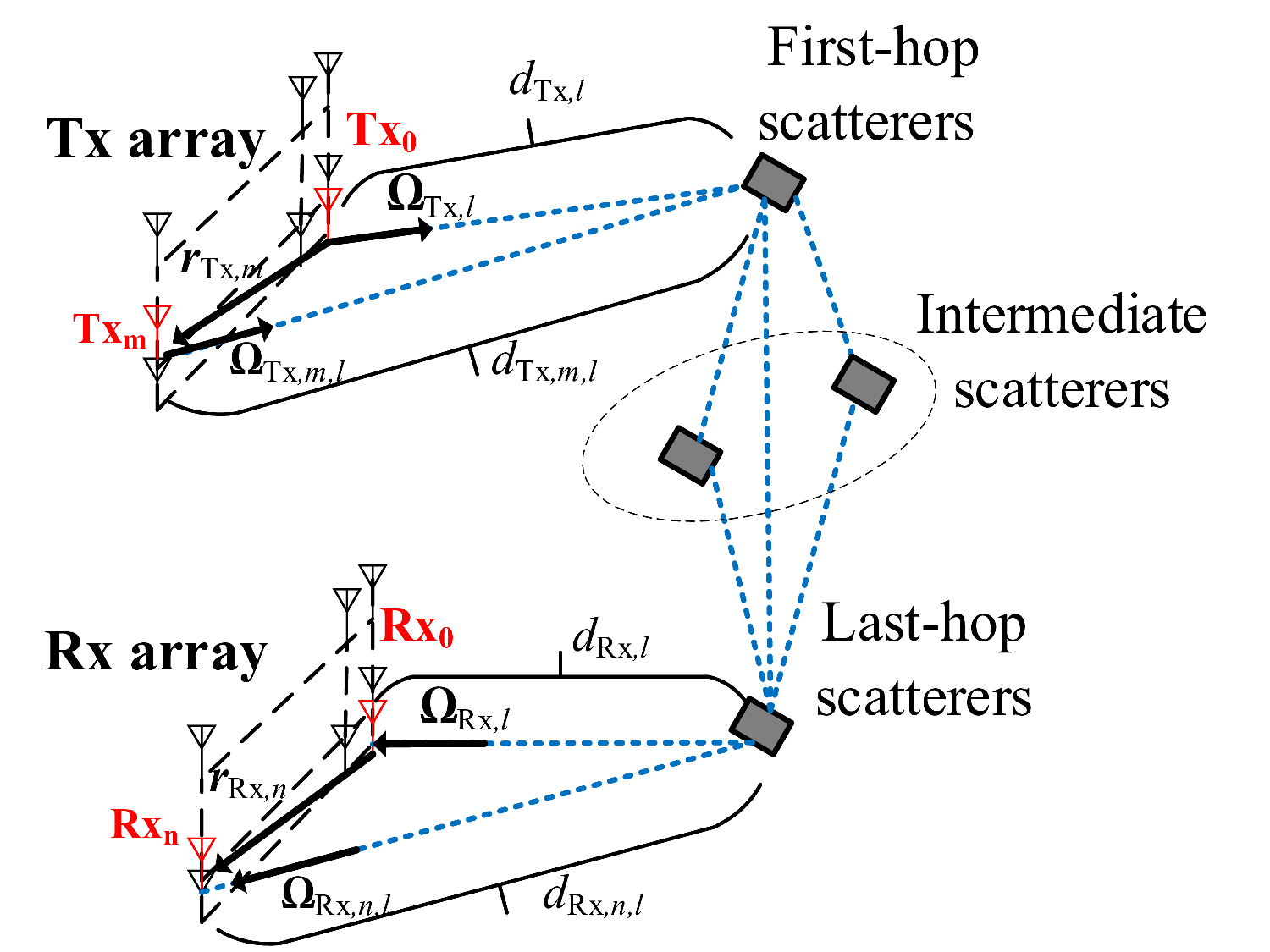}
           \vspace{-0.3cm}
    \caption{Illustration of a unified model based on the spherical wavefront.  
    }\label{fig:near_field}
       \vspace{-0.5cm}
\end{figure}
%%%%%%%%%%%%%%%
\subsection{Spherical wavefront based channel model}\label{Sec:signal_model}
The work assumes an $M$-Tx and $N$-Rx sensing system, where a signal comprises $P$ non-overlapping sub-bands in each frame, with each sub-band occupying $f_s$~Hz. Further, a coherent processing interval (CPI) in Rx comprises $Q$ frames each of duration $T_b$~s. Thus, a CPI occupies a bandwidth of $P f_s$~Hz and a duration of $Q T_b$~s.
%
%This $B = P f_s$ bandwidth signal , where $P$ is the number of sub-bands in a frame and $f_s$ is the sub-bandwidth, and $Q$ frames, each of duration $T_b$ forming 
%
Let $\mathbf{Z} \in \mathbb{C}^{MN \times PQ}$ represent the baseband equivalent measurement channel; herein, the $m+(n-1)M$th row $\mathbf{Z}[m+(n-1)M,:]$ denotes the measurement channel from the $m$th Tx to the $n$th Rx, the $p+(q-1)P$th column $\mathbf{Z}[:,p+(q-1)P]$ denotes the measurement channel of the $p$th sub-band in the $q$th frame, with $m = 1, 2, ..., M$ and $n = 1, 2, ..., N$ denote the indices of Tx and Rx, respectively; $p = 1, 2, ..., P$ and $q= 1, 2, ..., Q$ denote the indices of the sub-band and frame, respectively. 
The ${m+nM, q+qP}$th element of ${\bf Z}$, $Z_{m+nM, q+qP}$, 
% represents the channel from the $m$th Tx to the $n$th Rx at the $p$th frequency point of the $q$th frame 
consists of $L$ multipaths and takes the form 
% \textcolor{blue}{(check here the sampling of frequency domain)}
% \begin{equation} 
\begin{align}\label{eq:ch_1}
         Z_{m+nM, p+qP} = & \sum_{l=1}^{L} \alpha_{l} \Delta\alpha_{m,n,l} F_\text{Rx}(f_c, \mathbf{\Omega}_{\text{Rx},n,l} ) F_\text{Tx}(f_c, \mathbf{\Omega}_{\text{Tx},m,l} ) \nonumber
        \\ & e^{-j2\pi f_p \tau_{m,n,l}} e^{j2\pi f_{D,l} q T_b}e^{j\phi_l} + w(p,q), 
\end{align} 
% \vspace{-0.15cm}
% \end{equation}
with $f_p = p f_s$.
% \footnote{Beam squint is an issue of angle estimation in wideband systems, wherein designing sub-band dependent steering vectors is an effective mitigation approach \cite{8882325}.We consider both the signal model and algorithm design of this paper based on each sub-band, hence the beam squint effects can be well eliminated.} 
The Doppler frequency of the $l$th path is $f_{D,l} = - (f_c +p f_s) v_l / c$,
% \begin{align} \label{eq:Fd_1}
%     f_{D,l} = - \dfrac{(f_c +p f_s) v_l }{c}, ~~~l = 1, 2, ..., L
% \end{align}
where $f_c$ is the carrier frequency, $v_l$ is the radial velocity of the $l$th path, $c$ is the speed of light. Further,
$\phi_l$  and 
$\alpha_l$ are the constant phase and the propagation attenuation associated with the $l$th path of the reference channel, 
$\mathbf{\Omega}_{\text{Rx},n,l}$ denotes unit orientation vector of the AoA of the $l$th path at the $n$th Rx antenna, $\mathbf{\Omega}_{\text{Tx},m,l}$ denotes the unit orientation vector of the AoD of the $l$th path at the $m$th Tx antenna,
$F_\text{Rx}(f_c, \mathbf{\Omega}_{\text{Rx},n,l} )$ and $F_\text{Tx}(f_c, \mathbf{\Omega}_{\text{Tx},m,l} )$ denote the antenna gain at the $n$th Rx and the $m$th Tx, respectively for the $l$th path, and 
% % direction $\mathbf{\Omega}_{\text{Rx},n,l}$, 
%  denotes the antenna gain of $l$th path at ,
% % direction $\mathbf{\Omega}_{\text{Tx},m,l}$, 
$w(p,q)$ is the noise. %, $f_c$ is the carrier frequency, $v_l$ is the velocity of the $l$th path, $c$ is the speed of light, 
In \eqref{eq:ch_1}, $\Delta \alpha_{m,n,l}$ is the SNS amplitude between the $m$th-Tx $n$th-Rx channel and the reference channel of the $l$th path, and is modeled as $\Delta \alpha_{m,n,l} = {\tau_l }/{\tau_{m,n,l} }$ \cite{8713575},
% \begin{align} \label{eq:SNS_1}
%     \Delta \alpha_{m,n,l} = \dfrac{\tau_l }{\tau_{m,n,l} },
% \end{align}
% $\Delta \alpha_{\text{Tx},m,l}$ the amplitude non-stationarity between the $m$th Tx and the reference Tx of the $l$th path,
where $\tau_{m,n,l}$ denotes the propagation delay of the $l$th multipath between the $m$th Tx and the $n$th Rx calculated as $\tau_{m,n,l} = \tau_l + \Delta \tau_{\text{Rx},n,l} + \Delta \tau_{\text{Tx},m,l}$.
% \begin{equation} \label{eq:delay_1}
% %\begin{aligned}
%         \tau_{m,n,l} = \tau_l + \Delta \tau_{\text{Rx},n,l} + \Delta \tau_{\text{Tx},m,l}. 
% %\end{aligned} 
% \end{equation}
% In \eqref{eq:delay_1},  
$\tau_l$ denotes the delay of the reference channel, i.e., the delay between the reference $Tx_0$ via one or multiple scatterers to the reference $Rx_0$ as shown in Fig.~\ref{fig:near_field}, 
$\Delta \tau_{\text{Rx},n,l}$ and $\Delta \tau_{\text{Tx},m,l}$ represents 
the delay differences of the $n$th Rx and reference $Rx_0$ due to the last-hop scatterer, and the $m$th Tx and reference $Tx_0$ due to the first-hop scatterer of the $l$th MPC, respectively.
% the delay due to the spatial distance from the $n$th Rx to the reference Rx and from the $m$th Tx to the reference Tx of the $l$th MPC, respectively. 
In Fig.~\ref{fig:near_field}, using the unified distance representation, 
% where each MPC can be either a one-bounce or multiple-bounce path.
$\Delta \tau_{\text{Rx},n,l}$ and $\Delta \tau_{\text{Tx},m,l}$ can be represented as $\Delta \tau_{\text{Rx},n,l} = {d_{\text{Rx},n,l} -d_{\text{Rx},l}}/{c}$ and $\Delta \tau_{\text{Tx},m,l} = {d_{\text{Tx},m,l} -d_{\text{Tx},l}}/{c}$,
% \begin{equation} \label{eq:delay_2}
% \begin{aligned}
%         \Delta \tau_{\text{Rx},n,l} = \dfrac{d_{\text{Rx},n,l} -d_{\text{Rx},l}}{c}, \\
%         \Delta \tau_{\text{Tx},m,l} = \dfrac{d_{\text{Tx},m,l} -d_{\text{Tx},l}}{c},
% \end{aligned} 
% \end{equation}
where $d_{\text{Rx},n,l}$ and $d_{\text{Rx},l}$ denote the distance from the last-hop scatterer to the $n$th Rx and reference Rx, respectively, $d_{\text{Tx},m,l}$ and $d_{\text{Tx},l}$ denotes the distance from the first-hop scatterer to the $m$th Tx and reference Tx, respectively.
Further, the distance terms can be calculated using spherical coordinate representation as $ d_{\text{Rx},n,l} = \| d_{\text{Rx},l} \mathbf{\Omega}_{\text{Rx},l} - (\mathbf{r}_{\text{Rx},n}  -\mathbf{r}_\text{Rx}   )  \|$ and $d_{\text{Tx},m,l} = \| d_{\text{Tx},l} \mathbf{\Omega}_{\text{Tx},l} - (\mathbf{r}_{\text{Tx},m}  -\mathbf{r}_\text{Tx}   )  \|$ \cite{7501567},
% \begin{equation} \label{eq:delay_4}
% \begin{aligned}
%         d_{\text{Rx},n,l} = \| d_{\text{Rx},l} \mathbf{\Omega}_{\text{Rx},l} - (\mathbf{r}_{\text{Rx},n}  -\mathbf{r}_\text{Rx}   )  \|, 
% \end{aligned} 
% \end{equation}
% \begin{equation} \label{eq:delay_5}
% \begin{aligned}
%         d_{\text{Tx},m,l} = \| d_{\text{Tx},l} \mathbf{\Omega}_{\text{Tx},l} - (\mathbf{r}_{\text{Tx},m}  -\mathbf{r}_\text{Tx}   )  \|,
% \end{aligned} 
% \end{equation}
where $\mathbf{\Omega}_{\text{Rx},l}$ denote the AoA orientation vector from the last-hop scatterer to the reference Rx, $\mathbf{r}_{\text{Rx},n}$ and $\mathbf{r}_{\text{Rx}}$  denotes coordinates of the $n$th Rx antenna and the reference Rx antenna, respectively, 
$\mathbf{\Omega}_{\text{Tx},l}$ denotes the AoD orientation vector from the first-hop scatterer to the reference Tx, $\mathbf{r}_{\text{Tx},m}$ and $\mathbf{r}_{\text{Tx}}$  denotes coordinates of the $m$th Tx antenna and the reference Tx antenna, respectively. Kindly refer to Fig.~\ref{fig:near_field} for details.
The last constituents of \eqref{eq:ch_1} can be computed using
\vspace{-0.25cm}
\begin{align}\label{eq:delay_6}
        \mathbf{\Omega}_{\text{Rx},n,l} = \dfrac{d_{\text{Rx},l} \mathbf{\Omega}_{\text{Rx},l} - (\mathbf{r}_{\text{Rx},n}  -\mathbf{r}_\text{Rx}   )  }{d_{\text{Rx},n,l}}  ,  \\
        \label{eq:delay_7}
        \mathbf{\Omega}_{\text{Tx},m,l} = \dfrac{d_{\text{Tx},l} \mathbf{\Omega}_{\text{Tx},l} - (\mathbf{r}_{\text{Tx},m} -\mathbf{r}_{\text{Tx}}) }{d_{\text{Tx},m,l}}  ,
\end{align} 
% \begin{equation} 
% \begin{aligned}
%         \mathbf{\Omega}_{\text{Tx},m,l} = \dfrac{d_{\text{Tx},l} \mathbf{\Omega}_{\text{Tx},l} - (\mathbf{r}_{\text{Tx},m} -\mathbf{r}_{\text{Tx}}) }{d_{\text{Tx},m,l}}  , 
% \end{aligned} 
% \end{equation}
\begin{rem}
{\em In this context, the fundamental spherical wavefront offers a unified model for both near and far fields}. In \eqref{eq:delay_6}, when the reference distances, $d_{\text{Rx},l}$, are much larger than the antennas aperture, $\| \mathbf{r}_{\text{Rx},n}  -\mathbf{r}_\text{Rx} \|$, then $\mathbf{\Omega}_{\text{Rx},n,l} \approx \mathbf{\Omega}_{\text{Rx},l}$, i.e., the plane wavefront approximation of far-field holds. The same analysis applies to the Tx side in \eqref{eq:delay_7}.
\end{rem}

\subsection{Problem formulation and GM-SAGE solution}
Given the measurement channel $\mathbf{Z}$ in \eqref{eq:ch_1}, the aim is to estimate the scatterer parameters of the dictionary $\mathbf{\Psi}_k$\footnote{The coordinates, velocity, and the RCS of the scatterers parameterize the dictionary, hence we use the notation ${\mathbf{\Psi}_k}_{|{ \mathbf{\sigma}, \mathbf{v}, \mathbf{r}_\text{sc} }} $. The generation of a graph-based dictionary can refer to \cite{6275475}.},
% The dictionary $\mathbf{\Psi}$ is generated by the GT model, 
% i.e., the coordinates $\mathbf{r}_{\text{sc}}$, the reflectivity $\mathbf{\sigma}$, and the velocity $\mathbf{v}$ of vertices, 
the problem, i.e., the cost function is 
\vspace{-0.25cm}
\begin{equation} 
\begin{aligned} \label{eq:prob_1}
% \textcolor{red}{WX:Subscript\ of\ angles}\nonumber\\
    \underset{ \mathbf{\sigma}, \mathbf{v}, \mathbf{r}_\text{sc} }{\operatorname{\arg \min}}  \| \mathbf{Z} - \sum_{k=1}^{K} {\mathbf{\Psi}_k}_{|{ \hat{\mathbf{\sigma}}, \hat{\mathbf{v}}, \hat{\mathbf{r}}_\text{sc} }} \| ,
\end{aligned}
\end{equation}
where $K$ denotes the maximum bouncing in the channel. 
% where $\mathbf{Z}$ is measurement data, and $\mathbf{\Psi}$ is the above over-complete representation dictionary.
% The measurement matrix $\mathbf{Z}$ is a superposition of multi-bounce paths.  
% contains one-bounce, two-bounce, ..., and even $k$-bounce paths. 
% \subsection{GM-SAGE Solution}

The typical way of optimizing the $k$th-bounce paths is fixing the other $\{K-1\}$ multi-bounce paths, as $ \{ \hat{\mathbf{\sigma}}_k, \hat{\mathbf{v}}_k, \hat{\mathbf{r}}_{k,\text{sc}} \} =  \underset{ \mathbf{\sigma}, \mathbf{v}, \mathbf{r}_\text{sc} }{\operatorname{\arg \min}}  \| \mathbf{Z} - \sum_{k' \neq k}^{K} {\hat{\mathbf{\Psi}}_{k'}}  - \mathbf{\Psi}_{k|{ \hat{\mathbf{\sigma}}_k, \hat{\mathbf{v}}_k, \hat{\mathbf{r}}_{k,\text{sc}} }}  \|$, 
% \begin{equation}  \label{eq:prob_1_1}
%     \{ \hat{\mathbf{\sigma}}_k, \hat{\mathbf{v}}_k, \hat{\mathbf{r}}_{k,\text{sc}} \} =  \underset{ \mathbf{\sigma}, \mathbf{v}, \mathbf{r}_\text{sc} }{\operatorname{\arg \min}}  \| \mathbf{Z} - \sum_{k' \neq k}^{K} {\hat{\mathbf{\Psi}}_{k'}}  - \mathbf{\Psi}_{k|{ \hat{\mathbf{\sigma}}_k, \hat{\mathbf{v}}_k, \hat{\mathbf{r}}_{k,\text{sc}} }}  \| , 
%     % \\
%     % \hat{\mathbf{H}}_{k} =    \hat{\mathbf{H}}_{k-1} - \mathbf{\Psi}_{k|{ \hat{\mathbf{\sigma}}_k, \hat{\mathbf{v}}_k, \hat{\mathbf{r}}_{k,\text{sc}} }}   ,
% \end{equation}
where $\hat{\mathbf{\Psi}}_{k'}$ is the estimated channel of the $k'$-bounce paths, $\hat{\mathbf{\sigma}}_k, \hat{\mathbf{v}}_k, \hat{\mathbf{r}}_{k,\text{sc}}$ are reflectivity, velocity, and coordinates consisting of the $k$-bounce channels, respectively. 

Defining $\mathbf{z_1}$, $\mathbf{z_2}$, and $\mathbf{z_3}$ to be vectorized one-bounce, two-bounce and high-bounce channels in $\mathbf{Z}$, 
with the corresponding parameter sets $\mathbf{\theta}_1$, $\mathbf{\theta}_2$, and $\mathbf{\theta}_3$; $\mathbf{\psi_1}$ and $\mathbf{\psi_2}$ to be vectorized one-bounce and two-bounce dictionary, respectively. 
The problem in \eqref{eq:prob_1} can also written in the vector form as an iterative estimation comprising the following steps
\vspace{-0.15cm}
\begin{equation} 
\begin{cases} \label{eq:prob_1_3}
    { \hat{\mathbf{\theta}}_1}  =  \underset{ \mathbf{\theta}_1 }{\operatorname{\arg \min}}  \| \mathbf{z} - \mathbf{\psi}_{2}|{ \hat{\mathbf{\theta}}_2} - \mathbf{z}_3|{ \hat{\mathbf{\theta}}_3} - \mathbf{\psi}_{1}|{ {\mathbf{\theta}}_1}  \| , \\
    % \hat{\mathbf{h}}_{1} =     \mathbf{z}  - \mathbf{\psi}_{1}|{ \hat{\mathbf{\theta}}_1}   , \\
    { \hat{\mathbf{\theta}}_2} =  \underset{ \mathbf{\theta}_2 }{\operatorname{\arg \min}}  \| \mathbf{z} - \mathbf{\psi}_{1}|{ \hat{\mathbf{\theta}}_1} - \mathbf{z}_3|{ \hat{\mathbf{\theta}}_3} - \mathbf{\psi}_{2}|{ {\mathbf{\theta}}_2} \| ,  \\
    % \hat{\mathbf{h}}_{2} =      \hat{\mathbf{h}}_1 - \mathbf{\psi}_{2}|{ \hat{\mathbf{\theta}}_2}   ,  \\
    { \hat{\mathbf{\theta}}_3} =  \underset{ \mathbf{\theta}_3 }{\operatorname{\arg \min}}  \| \mathbf{z} - \mathbf{\psi}_{1}|{ \hat{\mathbf{\theta}}_1}  -  \mathbf{\psi}_{2}|{ \hat{\mathbf{\theta}}_2}- \mathbf{z}_3|{ {\mathbf{\theta}}_3} \| .
\end{cases} 
\end{equation}
% \vspace{-0.15cm}
{The recursive expectation and maximization step}
% For $l_1 = 1, 2, ..., L$, the $i$th iteration of the two-fold EM framework can be summarized as follows: 
With the initialization ${\mathbf{\theta}_1}^{(0)}$, ${\mathbf{\theta}_2}^{(0)}$, ${\mathbf{\theta}_3}^{(0)}$, the $i$th iteration of the optimization problem \eqref{eq:prob_1_3} is 
\vspace{-0.15cm}
\begin{equation} 
\begin{cases} \label{eq:prob_1_4}
    { \hat{\mathbf{\theta}}_1^{(i)}}  =  \underset{ \mathbf{\theta}_1 }{\operatorname{\arg \min}}  \| {\hat{\mathbf{z}}_1}^{(i)}  - \mathbf{\psi}_{1}|{ {\mathbf{\theta}}_1}  \| , \\
    % \hat{\mathbf{h}}_{1}^{(i)} =     \mathbf{z}  - \mathbf{\psi}_{1}|{ \hat{\mathbf{\theta}}_1^{(i)}}    , \\
    { {\hat{\mathbf{\theta}}_2}^{(i)}} =  \underset{ \mathbf{\theta}_2 }{\operatorname{\arg \min}}  \| {\hat{\mathbf{z}}_2}^{(i)} - \mathbf{\psi}_{2}|{ {\mathbf{\theta}}_2} \| ,  \\
    { {\hat{\mathbf{\theta}}_3}}^{(i)} =  \underset{ \mathbf{\theta}_3 }{\operatorname{\arg \min}}  \| \hat{\mathbf{h}}_2^{(i)} - \mathbf{z}_3|{ {{\mathbf{\theta}}_3}} \| ,
\end{cases} 
\end{equation}
% \vspace{-0.15cm}
with ${\hat{\mathbf{z}}_1}^{(i)}  =  \mathbf{z} - {\hat{\mathbf{z}}_2}^{(i-1)} - \hat{\mathbf{h}}_{2}^{(i-1)}$, ${\hat{\mathbf{z}}_2}^{(i)}  =  \mathbf{z} - {\hat{\mathbf{z}}_1}^{(i)} - \hat{\mathbf{h}}_{2}^{(i-1)}$, and $\hat{\mathbf{h}}_{2}^{(i)} =   \mathbf{z} - {\hat{\mathbf{z}}_1}^{(i)}  - {\hat{\mathbf{z}}_2}^{(i)}$.
% \begin{equation} 
% \begin{cases} \label{eq:prob_1_4_1}
%     {\hat{\mathbf{z}}_1}^{(i)}  =  \mathbf{z} - {\hat{\mathbf{z}}_2}^{(i-1)} - \hat{\mathbf{h}}_{2}^{(i-1)}, \\
%     {\hat{\mathbf{z}}_2}^{(i)}  =  \mathbf{z} - {\hat{\mathbf{z}}_1}^{(i)} - \hat{\mathbf{h}}_{2}^{(i-1)}, \\
%     \hat{\mathbf{h}}_{2}^{(i)} =   \mathbf{z} - {\hat{\mathbf{z}}_1}^{(i)}  - {\hat{\mathbf{z}}_2}^{(i)}.  
% \end{cases} 
% \end{equation}
% where $\hat{\mathbf{h}}_2$ is the vectorized residual channel vector of high-bounce paths in $\mathbf{z}$.
We can perform a SAGE iteration to solve each sub-problem in \eqref{eq:prob_1_4}. In total, the GM-SAGE contains three loops of the conventional SAGE algorithm. 
\begin{figure}[t]
    \vspace{-0.3cm}
     \centering
          \subfigure[Picture of the room. ]{\includegraphics[width=0.2\textwidth]{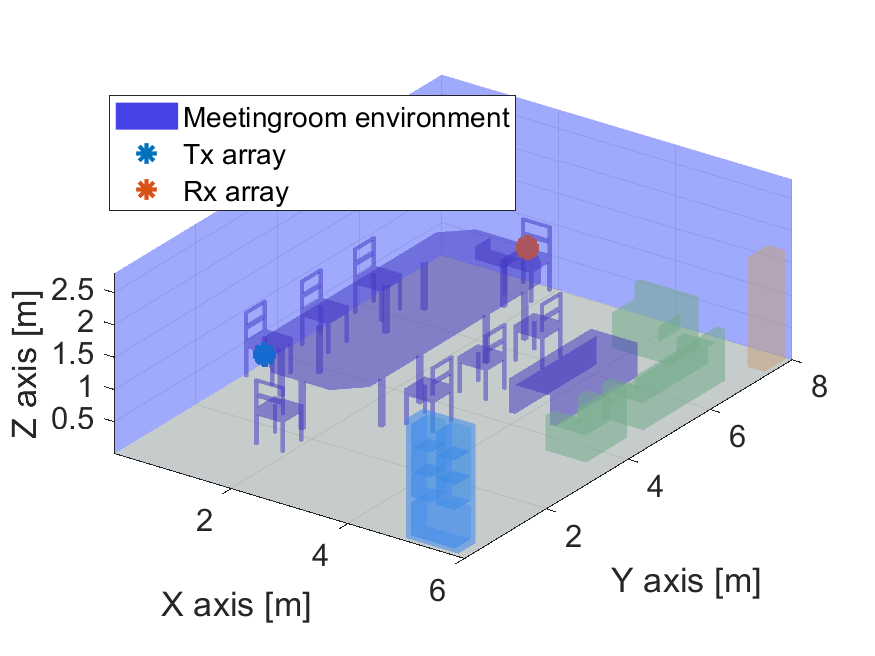}}
                        \hfill
    \subfigure[Three-bounce path tracks  ]{\includegraphics[width=0.22\textwidth]{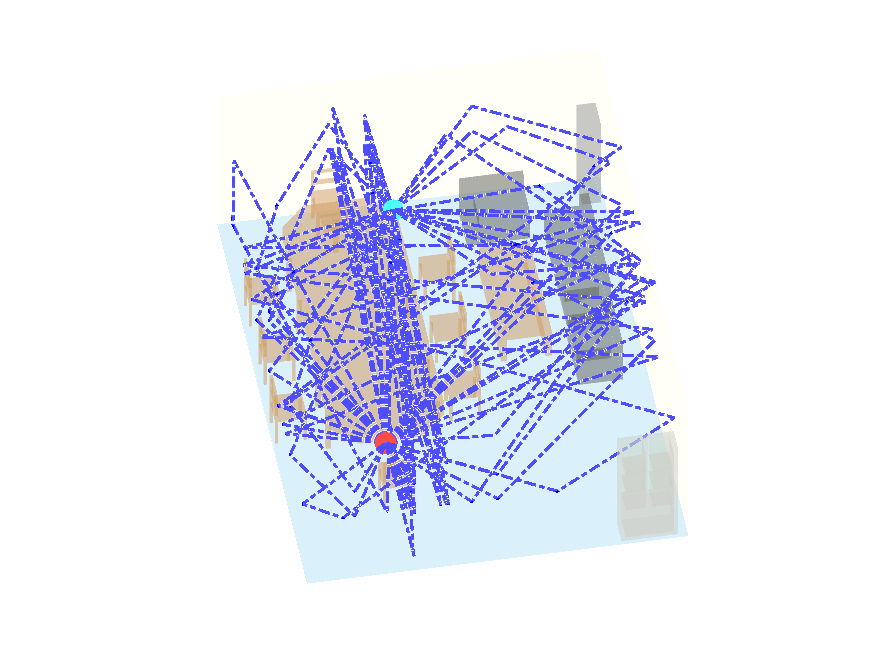}}
    \hfill
    \subfigure[The proposed GM-SAGE]{\includegraphics[width=0.235\textwidth]{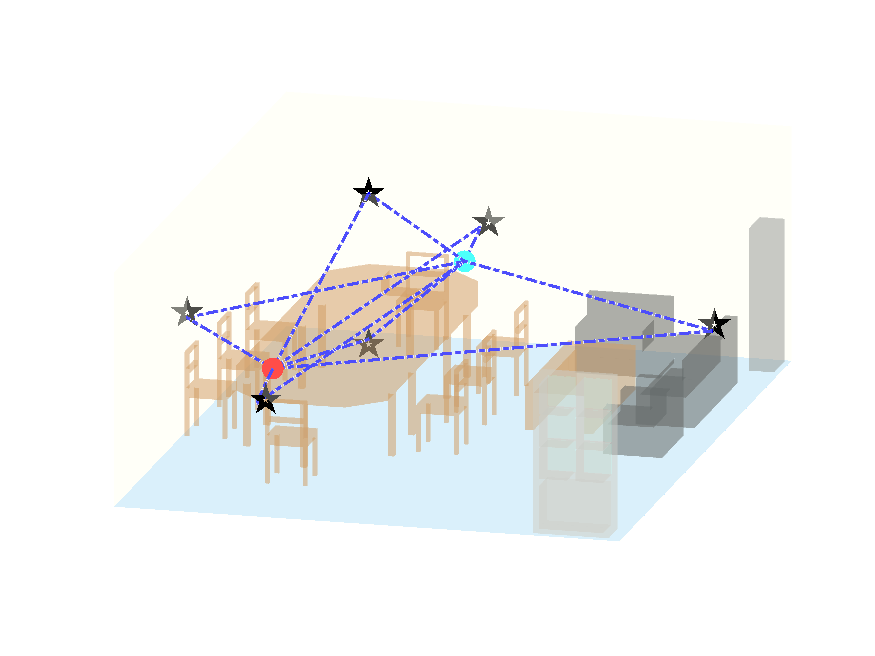}}
                        \hfill
    \subfigure[Methods considering one-bounce]{\includegraphics[width=0.235\textwidth]{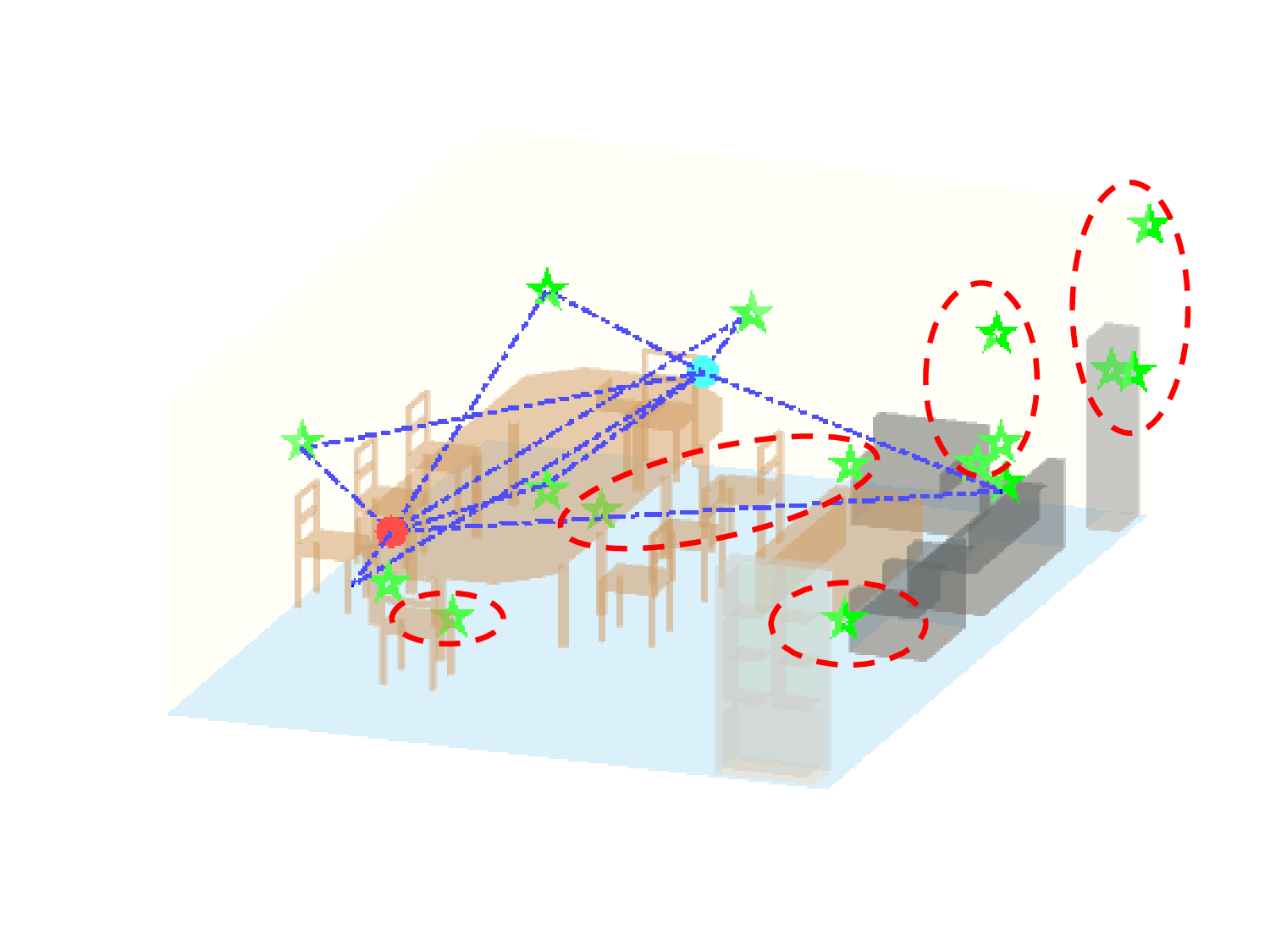}  }
    % \subfigure[$49 \times 49$ channel data]{\includegraphics[width=0.31\textwidth]{figures/C_PDP_three_bounce.eps}}
    % \hfill
     \vspace{-0.2cm}
        \caption{(1): considered scenario, (b): illustration of the multi-bounce multipath propagation, (c) and (d): scatterers localization and environment mapping. }
    \label{fig:simu_paths}
    \vspace{-0.3cm}
\end{figure}

\vspace{-0.3cm}
\subsection{Simulation case}\label{vali_ray}
%% Validation in a realistic complex indoor environment with multi-bounce reflections
% \subsubsection{Simulation settings and the multipath channel data}
The simulation scenario is a complex indoor office scenario as shown in Fig.~\ref{fig:simu_paths}~(a).
% where the coordinates of Tx and Rx are denoted by the blue and red stars, respectively. 
% The simulation parameters are shown in Tab.~\ref{tab:simu_configs_1}~column~\ref{vali_ray}.
%~\ref{tab:simu_configs_0} 
%%
% \begin{table}[t]
%     \caption{Bistatic simulation configurations used in Section~\ref{vali_ray}}
%     \centering
%     \begin{tabular}{lc}
%         \toprule
%         Configurations & Values \\
%         \midrule
%         % Central frequency $f_c$ [GHz] & $30$ \\
%         % Bandwidth $B$ [GHz] & 1 \\
%         % Frequency sweep interval $f_s$ [MHz] & $10$ \\
%         % Frequency step number $P$ & $101$ \\
%         SNR [dB] & $20$ \\  
%         Grid size [m] & 0.2 \\
%         Scenario scale [$m^3$] & $6 \times 8 \times 2.8 $ \\
%                 Array type & Rectangular \\
%         Number of Tx $M$ & $7 \times 7 = 49$ \\
%         Number of Rx $N$ & $7 \times 7 = 49$ \\
%         % Antenna distance [$m$]  & $ 0.5 \lambda $ \\
%         Antenna distance [$m$]  & $ 0.015 $ \\
%         Location of reference Tx & $(1.5, 1.5, 1.35)$ \\
%         Location of reference Rx & $(2.5, 6.5, 1.35)$ \\    
%         \bottomrule
%     \end{tabular}
%     \label{tab:simu_configs}
% \end{table}
The multi-bounce channel is generated via an open-access measurement-calibrated RT platform \cite{8438326}, where both the Tx and Rx array are using $7 \times 7$ antennas and the operation central frequency is $30$ GHz.
The $3D$ view of the up-to-three-bounce paths is used for validation, which is shown in Fig.~\ref{fig:simu_paths}~(b).
% , where the blue dashed lines denote the propagation traces operated in the simulator.
%
% In this simulation, measurement-calibrated channel data with up to three arcs are used,
% where the concatenated CIR of the total $49 \times 49$ MIMO channels collected in the structure in \eqref{eq:ch_1} is shown in Fig.~\ref{fig:simu_paths}~(c). 
% \subsubsection{Reconstructed scatterers}
% Here we use a $7 \times 7$ rectangular array, the Rayleigh distance is around $1.6$ meters, hence the two-bounce paths in this far-field region are with ambiguity as discussed in Section~\ref{Sec:ambiguty_near_far}.
% Hence, only the one-bounce paths and scatterers are identified in this simulation. 
% The channel data contains reflection paths.
% Unlike scattering, the bouncing point of different MIMO channels varies in perfect mirror reflections. However, the position variation of the one-bounce paths is not significant compared to the search grid, therefore, the proposed algorithm can still estimate the reflection paths.
The estimation one-bounce results using GM-SAGE are shown in Fig.~\ref{fig:simu_paths}~(c), where the estimated scatterers are mapped with the exact propagation track as denoted by blue dashed lines superimposed on the office scenario.
% , e.g., the four walls, the ceiling, and the table.
For comparison, the benchmark method that models the entire channel that is made up of one-bounce paths would result in ghost results in indoor scenarios, as shown in Fig.~\ref{fig:simu_paths}~(d), where red dots indicate ghost results. 

\section{Conclusion}
This paper addresses the multi-bounce near-field effects in indoor channel estimation and environment sensing.
% Firstly, the ambiguity of using geometrical parameters for sensing scatterers is analyzed. 
% Specifically, one-bounce paths can be utilized in both near- and far-field scenarios; 
% the two-bounce paths can be used in near-field scenarios while resulting in range ambiguity in far-field;
% more than three bounce paths can not be directly used to localize scatterers. 
% By formulating the scatterer localization into vertices searching problem, 
Specifically, this paper proposes a GM-SAGE algorithm to estimate the multi-bounce channels and reconstruct the locations of scatterers.
GT model-based vertices searching can inherently solve the range-angle coupling problem in near-field sensing. The scatterers of one- and two-bounce paths are localized, while the higher-bounce paths are identified as residual, hence avoiding ghost results.
Finally, the proposed GM-SAGE is validated by RT simulations in complex indoor scenarios. 
% Besides, numerical simulation approves the ambiguity of two-bounce estimations using geometrical parameters in far-field scenarios. 
% In the two-bounce Monte Carlo simulations, we compare the near- and far-field ambiguity in two perspectives: (i) keep the distribution of scatterers constant, with the increase of the antenna aperture size, i.e., the Rayleigh distance, the localization performance improves; (ii) keep the number of antennas constant, the farther away the scatterer is from the near-field, the worse the positioning results.
% There are still challenges for future orientation, such as mirror reflection ambiguity and
% complicated SNS in radio-based sensing. 

{\em Acknowledgment:}
This work was supported by the Luxembourg National Research Fund (FNR) through the BRIDGES project MASTERS under grant BRIDGES2020/IS/15407066. The idea presented in this abstract has been extended to a journal paper \cite{Yuan_TWC25}.

% \newpage

\bibliographystyle{IEEEtran}
\bibliography{ref.bib}

% Generated by IEEEtran.bst, version: 1.14 (2015/08/26)
\begin{thebibliography}{10}
\providecommand{\url}[1]{#1}
\csname url@samestyle\endcsname
\providecommand{\newblock}{\relax}
\providecommand{\bibinfo}[2]{#2}
\providecommand{\BIBentrySTDinterwordspacing}{\spaceskip=0pt\relax}
\providecommand{\BIBentryALTinterwordstretchfactor}{4}
\providecommand{\BIBentryALTinterwordspacing}{\spaceskip=\fontdimen2\font plus
\BIBentryALTinterwordstretchfactor\fontdimen3\font minus \fontdimen4\font\relax}
\providecommand{\BIBforeignlanguage}[2]{{%
\expandafter\ifx\csname l@#1\endcsname\relax
\typeout{** WARNING: IEEEtran.bst: No hyphenation pattern has been}%
\typeout{** loaded for the language `#1'. Using the pattern for}%
\typeout{** the default language instead.}%
\else
\language=\csname l@#1\endcsname
\fi
#2}}
\providecommand{\BIBdecl}{\relax}
\BIBdecl

\bibitem{BalanisConstantineA2005At:a}
C.~A. Balanis, ``\BIBforeignlanguage{eng}{Antenna theory : analysis and design},'' New York, 2005.

\bibitem{zhi2024performance}
K.~Zhi, C.~Pan, H.~Ren, K.~K. Chai, C.-X. Wang, R.~Schober, and X.~You, ``Performance analysis and low-complexity design for xl-mimo with near-field spatial non-stationarities,'' \emph{IEEE Journal on Selected Areas in Communications}, 2024.

\bibitem{zhang2022scatterer}
G.~Zhang, X.~Cai, J.~{\O}. Nielsen, G.~F. Pedersen, and F.~Tufvesson, ``A scatterer localization method using large-scale antenna array systems,'' in \emph{2022 IEEE Conference on Antenna Measurements and Applications (CAMA)}.\hskip 1em plus 0.5em minus 0.4em\relax IEEE, 2022, pp. 1--4.

\bibitem{xing2021millimeter}
Y.~Xing, T.~S. Rappaport, and A.~Ghosh, ``Millimeter wave and sub-{THz} indoor radio propagation channel measurements, models, and comparisons in an office environment,'' \emph{IEEE Communications Letters}, vol.~25, no.~10, pp. 3151--3155, 2021.

\bibitem{10179246}
Z.~Zhou, C.-X. Wang, L.~Zhang, J.~Huang, L.~Xin, E.-H. Aggoune, and Y.~Miao, ``A novel {SAGE} algorithm for estimating parameters of wideband spatial nonstationary wireless channels with antenna polarization,'' \emph{IEEE Transactions on Antennas and Propagation}, vol.~71, no.~9, pp. 7457--7472, 2023.

\bibitem{10038714}
G.~Jing, J.~Hong, X.~Yin, J.~Rodriguez-Pineiro, and Z.~Yu, ``Measurement-based 3-d channel modeling with cluster-of-scatterers estimated under spherical-wave assumption,'' \emph{IEEE Transactions on Wireless Communications}, vol.~22, no.~9, pp. 5828--5843, 2023.

\bibitem{liu20223d}
Y.~Liu, L.~Wu, M.~Alaee-Kerahroodi \emph{et~al.}, ``A {3D} indoor localization approach based on spherical wave-front and channel spatial geometry,'' in \emph{2022 IEEE 12th Sensor Array and Multichannel Signal Processing Workshop (SAM)}.\hskip 1em plus 0.5em minus 0.4em\relax IEEE, 2022, pp. 101--105.

\bibitem{8713575}
X.~Cai and W.~Fan, ``A complexity-efficient high resolution propagation parameter estimation algorithm for ultra-wideband large-scale uniform circular array,'' \emph{IEEE Transactions on Communications}, vol.~67, no.~8, pp. 5862--5874, 2019.

\bibitem{ling2017experimental}
C.~Ling, X.~Yin, H.~Wang, and R.~S. Thom{\"a}, ``Experimental characterization and multipath cluster modeling for 13--17 ghz indoor propagation channels,'' \emph{IEEE Transactions on antennas and propagation}, vol.~65, no.~12, pp. 6549--6561, 2017.

\bibitem{de2023analysis}
M.~F. De~Guzman and K.~Haneda, ``Analysis of wave-interacting objects in indoor and outdoor environments at 142 ghz,'' \emph{IEEE Transactions on Antennas and Propagation}, 2023.

\bibitem{wang2021room}
Y.~Wang, K.~Ho, and L.~Huang, ``Room geometry estimation using the multipath delays,'' \emph{IEEE Signal Processing Letters}, vol.~28, pp. 1380--1384, 2021.

\bibitem{zhang20235g}
H.~Zhang, H.~Wymeersch, and F.~Wen, ``{5G NLOS} positioning with multi-bounce mitigation by iterative weighted least squares,'' in \emph{2023 IEEE Globecom Workshops (GC Wkshps)}.\hskip 1em plus 0.5em minus 0.4em\relax IEEE, 2023, pp. 92--97.

\bibitem{10556695}
E.~Rastorgueva-Foi, O.~Kaltiokallio, Y.~Ge, M.~Turunen, J.~Talvitie, B.~Tan, M.~Furkan~Keskin, H.~Wymeersch, and M.~Valkama, ``Millimeter-wave radio {SLAM}: End-to-end processing methods and experimental validation,'' \emph{IEEE Journal on Selected Areas in Communications}, pp. 1--1, 2024.

\bibitem{7501567}
J.~Chen, S.~Wang, and X.~Yin, ``A spherical-wavefront-based scatterer localization algorithm using large-scale antenna arrays,'' \emph{IEEE Communications Letters}, vol.~20, no.~9, pp. 1796--1799, 2016.

\bibitem{6275475}
T.~Pedersen, G.~Steinbock, and B.~H. Fleury, ``Modeling of reverberant radio channels using propagation graphs,'' \emph{IEEE Transactions on Antennas and Propagation}, vol.~60, no.~12, pp. 5978--5988, 2012.

\bibitem{8438326}
D.~He, B.~Ai, K.~Guan, L.~Wang, Z.~Zhong, and T.~Kurner, ``The design and applications of high-performance ray-tracing simulation platform for 5g and beyond wireless communications: A tutorial,'' \emph{IEEE Communications Surveys $\&$ Tutorials}, vol.~21, no.~1, pp. 10--27, 2019.

\bibitem{Yuan_TWC25}
Y.~Liu, L.~Wu, X.~Cai, and M.~R.~B. Shankar, ``Graph-based multi-bounce modeling and channel parameter estimation for indoor sensing,'' \emph{IEEE Transactions on Wireless Communications}, vol.~24, no.~5, pp. 4219--4234, 2025.

\end{thebibliography}
% \bibliography{references}
\end{document}